\documentclass[a4paper, 11pt, twoside]{article}
\usepackage{fullpage}  
\usepackage[lmargin=0.6in,rmargin=0.6in,tmargin=0.6in,headsep=.2in]{geometry}
\usepackage{graphicx}
\usepackage{forloop}
\usepackage{etoolbox}
\usepackage{adjustbox}
\usepackage{subcaption}
\usepackage{calc}
\usepackage{wrapfig}
\usepackage{booktabs} 
\usepackage{xtab} 
\usepackage[dvipsnames]{xcolor}
\usepackage[normalem]{ulem}
\usepackage{sectsty} 
\usepackage{hyperref}

\hypersetup{
    colorlinks=true,
    urlcolor=blue,
    linkcolor=black,
    citecolor=black
}

\usepackage{fancyhdr}
\usepackage{amsmath,amsfonts,amssymb,amsthm,epsfig,epstopdf,url,array}
\usepackage[retainorgcmds]{IEEEtrantools}
\usepackage{makeidx,lscape,pict2e}
\usepackage{upgreek}

\allsectionsfont{\color{Brown}}

\makeatletter
\def\@seccntformat#1{\@ifundefined{#1@cntformat}%
{\csname the#1\endcsname\;}
{\csname #1@cntformat\endcsname}
}
\def\section@cntformat{\thesection.\;} 
\def\subsection@cntformat{\thesubsection.\;} 
\makeatother

\fancypagestyle{first}{%
  \fancyhf{}
  \fancyhead[L]{\includegraphics[scale=0.8]{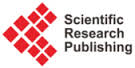}}%
  \fancyfoot[L]{{\footnotesize \textsf{DOI: \href{10.4236/jilsa.2024.162006}
  {\color{blue}\uline{10.4236/jilsa.2024.162006}} $\;$April 30, 2024}}}%
  \fancyhead[R]{{\bf\small \textsf{Journal off Intelligent Learning Systems and Applications, 2024, 16, 80-90}}\\
\href{http://www.scirp.org/journal/jilsa}
{\color{blue}\uline{\textsf{http://www.scirp.org/journal/jilsa}}}\\
\textsf{ISSN Online:: 2150-8410}\\
\textsf{ISSN Print:: 2150-8402}}%
}

\theoremstyle{definition}

\begin{document}
\thispagestyle{first}
\vspace*{3cm}
{\noindent\huge\bf Transfer learning approach to Classify the X-ray image that corresponds to corona disease Using ResNet50 pre-trained by ChexNet}\\[1cm]
{\bf\large Mahyar Bolhassani}\\[0.5cm]
University of Louisville, KY, USA\\
Email: mahyar.bolhassani@louisville.edu\\
\begin{wraptable}{l}{5.1cm}
{\footnotesize
\begin{xtabular*}{0.3\textwidth}{p{5cm}}
\noindent{\bf How to cite this paper:} Bolhassani, M. (2024) Transfer learning approach to Classify the X-ray image that corresponds to corona disease Using ResNet50 pre-trained by ChexNet, Journal of Intelligent Learning
Systems and Applications, {\bf 16},    80-90.\\ 

\url{http://dx.doi.org/10.4236/jilsa.2024.162006}\\
Accepted: April 27, 2024
Published: April 30, 2024
{\bf Received: February 29, 2024}\\
{\bf Accepted: April 27, 2024}\\
{\bf Published: April 30, 2024}\\
Copyright \copyright$\;$2024 by author(s) and Scientific Research Publishing Inc.\\
This work is licensed under the Creative Commons Attribution International License (CC BY 4.0).\\
\url{http://creativecommons.org/licenses/by/4.0/}\\
\end{xtabular*}
}
\end{wraptable}
{\color{Brown}\rule{0.7\textwidth}{2pt}}\\[0.2cm]
{\color{Brown}\bf\large Abstract}\\
The COVID-19 pandemic has had a widespread negative impact globally. It shares symptoms with other respiratory illnesses such as pneumonia and influenza, making rapid and accurate diagnosis essential to treat individuals and halt further transmission. X-ray imaging of the lungs is one of the most reliable diagnostic tools. Utilizing deep learning, we can train models to recognize the signs of infection, thus aiding in the identification of COVID-19 cases. For our project, we developed a deep learning model utilizing the ResNet50 architecture, pre-trained with ImageNet and CheXNet datasets. We tackled the challenge of an imbalanced dataset, the CoronaHack Chest X-Ray dataset provided by Kaggle, through both binary and multi-class classification approaches. Additionally, we evaluated the performance impact of using Focal loss versus Cross-entropy loss in our model.
\vspace{0.5cm}\\
{\color{Brown}\bf\large Keywords}\\
X-ray Classification; Convolutional Neural Network; ResNet; Transfer Learning; Supervised Learning, COVID-19, Chest X-ray.
\vspace{0cm}\\
{\color{Brown}\rule{0.7\textwidth}{2pt}}

\section{Introduction}
The World Health Organization (WHO) labeled the COVID-19 outbreak as a pandemic in January 2020, a crisis that wrought havoc across various sectors such as the economy, politics, and education worldwide. In light of its high transmissibility, immediate measures were prioritized to slow the spread of the virus. A vital part of this response required a dependable, rapid, and widely accessible diagnostic tool, for which medical image processing through deep learning emerged as a solution. This approach is particularly useful because one of the primary indicators of COVID-19 is lung infection, visible on X-ray images. Despite presenting symptoms similar to other respiratory conditions like pneumonia and influenza, X-ray imaging offers a reliable means of differentiation. Consequently, throughout the pandemic, numerous organizations have amassed a collection of X-ray images ranging from healthy to COVID-19-affected lungs.

In our research, we first addressed binary classification to differentiate bacterial from viral infections, prompted by the imbalanced nature of existing datasets, using a transfer learning approach. Following this, we extended the technique to a four-category classification system. To mitigate the issue of dataset imbalance, we employed Focal loss, aimed at offsetting the disparity in data class representation. Building on our findings, we sought to enhance the model's accuracy by generating synthetic input data via conditional Generative Adversarial Networks (GANs).

\section{Materials and Methods}
Automatic classification of X-ray scan images is a challenging task when we have a highly imbalanced dataset. Therefore, in this article, we are trying to find the best approach to increase the classification accuracy.

\subsection{Benchmark Datasets}
In our research, we utilized the CoronaHack Chest X-Ray Dataset provided by the competition organizers to train a deep learning model for the classification of X-ray images indicating COVID-19 infection \cite{Kaggle}. The dataset is divided into four primary groups: Normal, Bacteria, Virus, and COVID-19. The distribution of the dataset is notably uneven, containing 1,575 normal cases, 2,778 bacterial cases, 1,494 viral cases, and just 82 COVID-19 cases. To visualize this disparity, we created a class distribution chart that clearly depicts the dataset's imbalance. Additionally, we included a histogram that outlines the allocation of training and testing samples as designated by the competition's organizers, which is depicted in Figure 1.
\begin{figure}[!htbp]
    \centering
    \hspace{20mm}
    \includegraphics[width=10.5 cm]{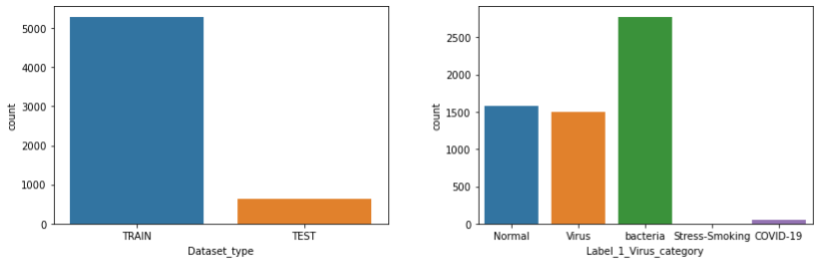}
    \caption{On the left: train-test distribution, on the right: Class distribution of a dataset.\label{fig1}}
\end{figure}

\begin{figure}[!htbp]
    \centering
    \begin{adjustbox}{addcode={\begin{minipage}{\width}}{\caption{%
        Histogram distribution of data samples.
        }\end{minipage}},right}
        \includegraphics[width=0.7\textwidth]{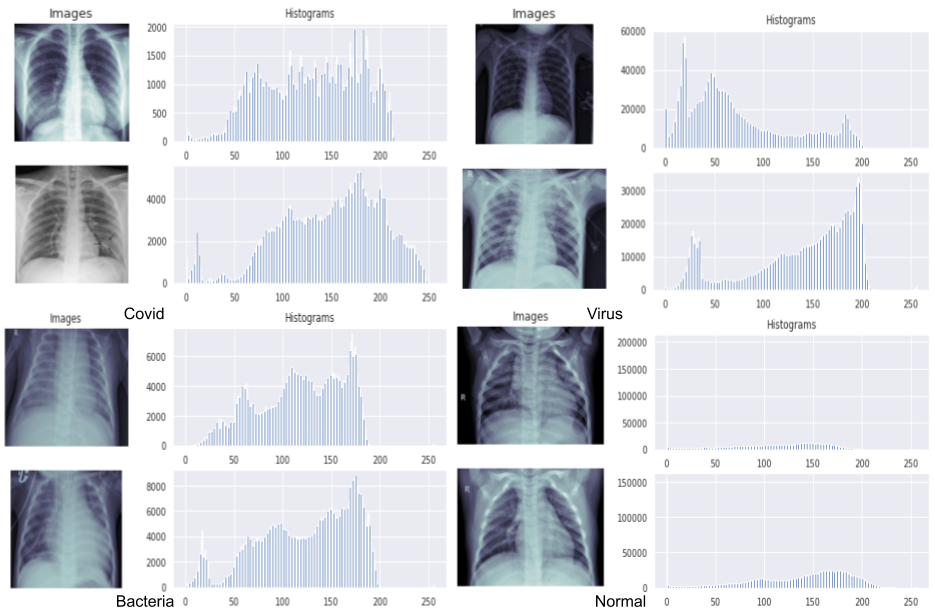}
    \end{adjustbox}
    \label{fig2}
\end{figure}

To gain a deeper insight into the dataset, plotting the data provides a clearer perspective on how each category differs from the others. Figure \ref{fig2} displays two sample images from each of the four classes, alongside a histogram that represents the distribution of intensity values for each class.
Observations from Figure 2 indicate distinct differences in the intensity distributions among samples from different classes. This variation serves as a promising indicator that a deep learning model could be effectively trained to distinguish between them.

\subsection{Data augmentation}
Deep learning models require a substantial quantity of labeled data to make precise predictions on test sets. However, acquiring annotated medical imagery is challenging, costly, and time-intensive, necessitating the involvement of medical professionals to annotate or label the images \cite{thesis}. In our context, this means needing specialists to categorize lung X-ray images across a spectrum from normal to those indicating COVID-19 infection. To address the limited availability of training samples, we must devise an effective strategy.

Data augmentation is one approach to mitigate the issue of limited data. This technique allows us to enhance the model's performance by introducing variety in the training data, enabling the model to recognize samples with various alterations. In our study, we have implemented numerous augmentation strategies, such as random horizontal and vertical flips with a 50\% probability, and random rotations with a 30\% likelihood.

\subsection{Architecture}
To train a model capable of classifying lung X-ray images, we have selected two architectures recognized for their efficacy in medical image processing: DenseNet and ResNet, both of which have demonstrated strong performance with medical datasets.

\subsubsection{ResNet}
\begin{figure}[!htbp]
    \centering
    \begin{adjustbox}{addcode={\begin{minipage}{\width}}{\caption{%
        On the left: DenseNet \cite{DenseNet}, on the right: ResNet \cite{ResNet}.
        }\end{minipage}},right}
        \includegraphics[width=0.7\textwidth]{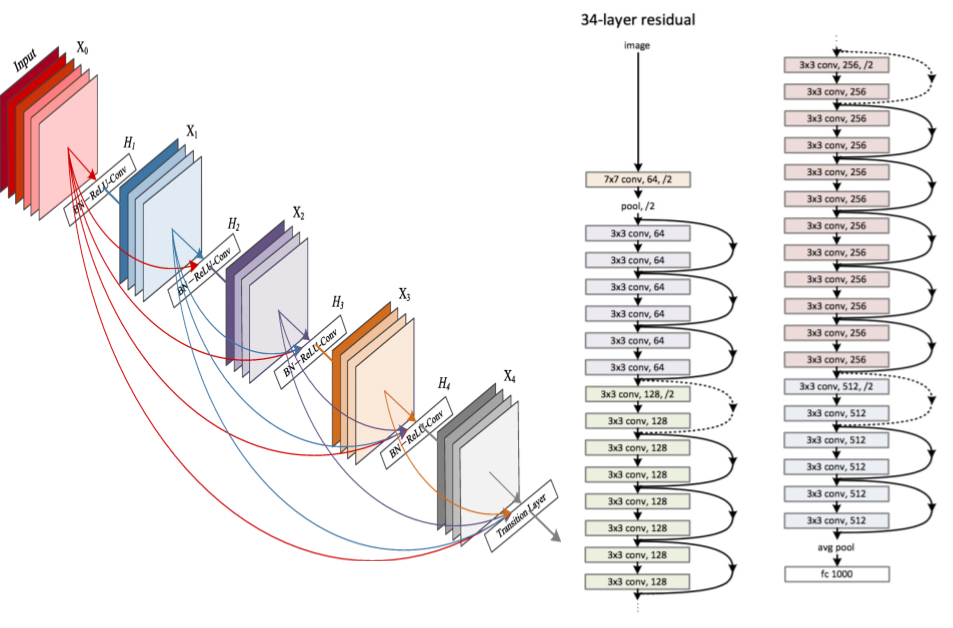}
    \end{adjustbox}
    \label{fig3}
\end{figure}

Developed to address the issue of vanishing gradients in deep convolutional networks, ResNet introduces a method for preserving the gradient by using skip connections between blocks. This allows the gradient to have a shortcut path, maintaining its strength throughout the network's depth. A detailed diagram of the ResNet architecture \cite{ResNet} is provided in figure \ref{fig3}.
The ResNet50 architecture is a deep learning model designed for image recognition, part of the ResNet (Residual Network) family that introduced the concept of residual learning to ease the training of very deep networks. It comprises 50 layers, including convolutional layers, activation layers (ReLU), batch normalization layers, and pooling layers, structured around residual blocks. These blocks have skip connections that allow inputs to bypass one or two layers and be added back to the output of a layer, combating the vanishing gradient problem in deep networks. The model is initialized with specific parameter settings, such as filters of varying sizes (e.g., 7x7 in the first convolutional layer, followed by 3x3 and 1x1 in subsequent layers) and strides to control the convolutional step. Training ResNet50 involves using a large dataset (e.g., ImageNet) with a backpropagation algorithm, typically employing techniques like stochastic gradient descent, with momentum and weight decay for optimization. The model also uses a softmax function in the output layer for classification. ResNet50's design enables it to learn robust feature representations for a wide variety of images, achieving remarkable accuracy in image classification tasks.

\subsubsection{DenseNet}
The traditional convolutional neural network has a problem in a case when we use a much deeper network. In this case, since the path of information from input to output becomes very long (as a result of a deep network) for both forward and backward paths, it is likely to face a vanishing gradient. DenseNet \cite{DenseNet} were introduced to solve the issue mentioned above. In this architecture, the output of each layer is passed to the input of all other layers. This way, not only does the gradient vanishing solve but also we need to manage fewer parameters in comparison to the vanilla CNN. More details of DenseNet architecture are shown in Figure \ref{fig3} on the left image.

\subsection{Loss function}
The next step is defining our loss function for the training part. We know that choosing the right loss function which is a hyperparameter depends on the problem we are facing. Therefore, to tackle a multi-classification problem multi-class cross entropy loss (Categorical Cross Entropy Loss Function) seems a wise choice. Equation (1) is the formula for this loss function. 
\begin{align}
CE = -\sum_{i=1}^{C'=2}t_{i} \log (s_{i}) = -t_{1} \log(s_{1}) - (1 - t_{1}) \log(1 - s_{1}) 
\end{align} 

Looking deeper into the problem we have, something forces us to be more cautious and it is the imbalanced data samples that we have as the input of our model. Focal loss \cite{focaloss} is another choice that we can leverage its properties to enhance the performance of our model. This loss function tries to generate a class weighting system in order to balance the samples in each batch size of data. Equation (2) shows the details regarding this function.
\begin{align}
FL (pt) = -\alpha t (1-pt) ^ \gamma \log(pt)
\end{align} 

\subsection{Metric}
One of the most important steps in the deep learning process is to define metrics which means how we evaluate the performance of our model. We chose to use accuracy which is the simplest metric. Accuracy is defined by finding the number of predictions per class divided by the number of predictions on each epoch. The average accuracy that we report in this project is obtained by averaging the accuracy of each epoch.
 
\subsection{Weighted class}
Our dataset is not balanced so we are not sure whether in each mini batch of data, all class samples exist or not. Therefore, one of our solutions for the mentioned problem is to use the weightedRandomSampler function in Pytorch. We first calculated the number of samples in each class. Then, we got the reciprocal value of them and considered them as weights of each class. In the next step, we passed the class weights to each class label.

\section{Detail implementation}
Transfer learning is our dominant approach to classifying COVID-19-infected cases. According to the fact that we have a small and imbalanced dataset in which covid-19 class is considerably smaller than other classes. This issue can deteriorate our prediction, as a result, we need to seek a method to alleviate the unbalanced class distribution of the classes. 

We started with ResNet50 architecture pre-trained by the ImageNet dataset. Then, we applied a weight sampling approach to have an equal number of samples from each class. We, also, examine using both categorical cross entropy and focal loss to see the effect of focal loss on our imbalanced dataset. Although our attempts paid off and the accuracy on both training and validation increased, still we tried to improve the accuracy of the model. We know that medical images are different from natural images, for instance, medical images are mostly in grayscale. Hence, a question arises here whether using a model pre-trained by ImageNet which is a collection of natural images, is a wise choice or not. This question motivated us to research this issue so we ended up to \cite{covidAID} in which they trained their model using CheXNet. CheXNet \cite{chexnet} is a model based on DenseNet architecture that is trained on large brain MRI scan images. We decided to use the same approach to check if it improved the performance of our model or not. Finally, we added some COVID-19 images to the dataset (our dataset has a very limited number of COVID-19 samples).

For training the model, we chose an epoch number of 20 for time and GPU restrictions. Also, we used the Adam optimizer with an initial learning rate with a value of 0.001. The learning rate is updated every 10 epochs and multiplied with a factor of 0.5 to avoid vanishing or exploding gradient.

\section{Experimental results}
To show our results, we divided each selection of hyperparameters into a section. This will allow us to explain the output and drawbacks of each in detail. We considered Adam optimizer with the learning rate scheduler explained in section 3. In all figures in this section, the images on the left represent Training and validation accuracy vs the number of epochs, while the images on the right illustrate Training and validation loss vs the number of epochs. 

The experimental results of each subsection are gathered in Table \ref{tab:1}.

\begin{table}[htbp]
  \begin{flushright} 
    \caption{Accuracy results for each proposed model}
    \begin{tabular}{ccccccc}
    \hline
    Models & PRCE & PRCW & PRFL & PDCXCE & PDCXFL & RCE \\
    \hline
    Tr Acc. & 81.06  & 86.06 & 85.84 & 85.28 & 87.95 & 81.608 \\
    Val Acc. & 65.39  & 73.43 & 75.0 & 76.062 & 73.87 & 69.93 \\
    Tst Acc. & 62.1  & 80.2 & 75.5 & 80.0 & 75.56 & 71.11 \\
    \hline
    \end{tabular}%
    \label{tab:1}%
  \end{flushright}
\end{table}%

According to the results provided in Table \ref{tab:1}, pretraining DenseNet on the CheXNet dataset outperforms other models.

\subsection{ResNet pre-trained on ImageNet}
\subsubsection{Considering CE loss (PRCE)}
In this section, we train our model based on ResNet50 pre-trained by ImageNet. The loss function is categorical cross-entropy. 

\subsubsection{Considering CE loss and weighted-class (PRCEW)}
In this section, we train our model based on ResNet50 pre-trained by ImageNet. The loss function is categorical cross entropy while including weighted classes in the model to decrease the adverse effects of imbalanced data on the results. 

\subsubsection{Considering FL loss (PRFL)}
In this section, we train our model based on ResNet50 pre-trained by ImageNet. The loss function is the focal loss to reduce the imbalanced distribution of sample data. 

\begin{figure}[!htbp]
    \centering
    \begin{adjustbox}{addcode={\begin{minipage}{\width}}{\caption{%
        First row: ResNet pre-trained model with CE loss, Second row: ResNet pre-trained model with weighted class CE loss, Third row: ResNet pre-trained model with FL loss.
        }\end{minipage}},right}
        \includegraphics[width=0.7\textwidth]{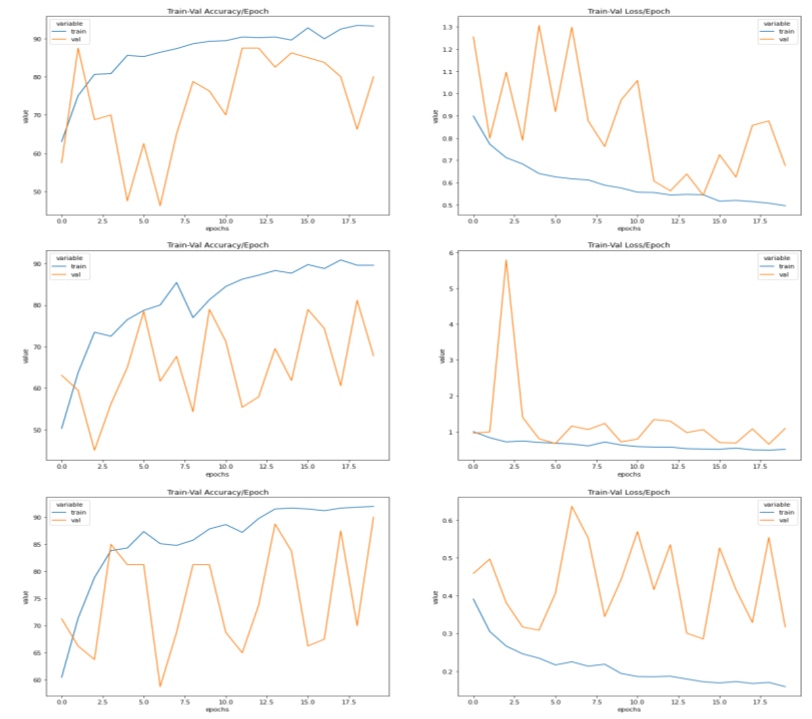}
    \end{adjustbox}
    \label{fig4}
\end{figure}

\subsection{DenseNet pre-trained on CheXNet}
\subsubsection{Considering CE loss (PDCXCE)}
In this section, we train our model based on DenseNet121 pretrained on CheXNet considering CE loss. 

\subsubsection{Considering FL loss(PDCXFL)}
In this section, we train our model based on DenseNet121 pretrained by CheXNet considering FL loss. 

\begin{figure}[!htbp]
    \centering
    \begin{adjustbox}{addcode={\begin{minipage}{\width}}{\caption{%
        First row: DenseNet pretrained on CheXNet with CE, Second row: DenseNet pretrained on CheXNet with FL
        }\end{minipage}},right}
        \includegraphics[width=0.7\textwidth]{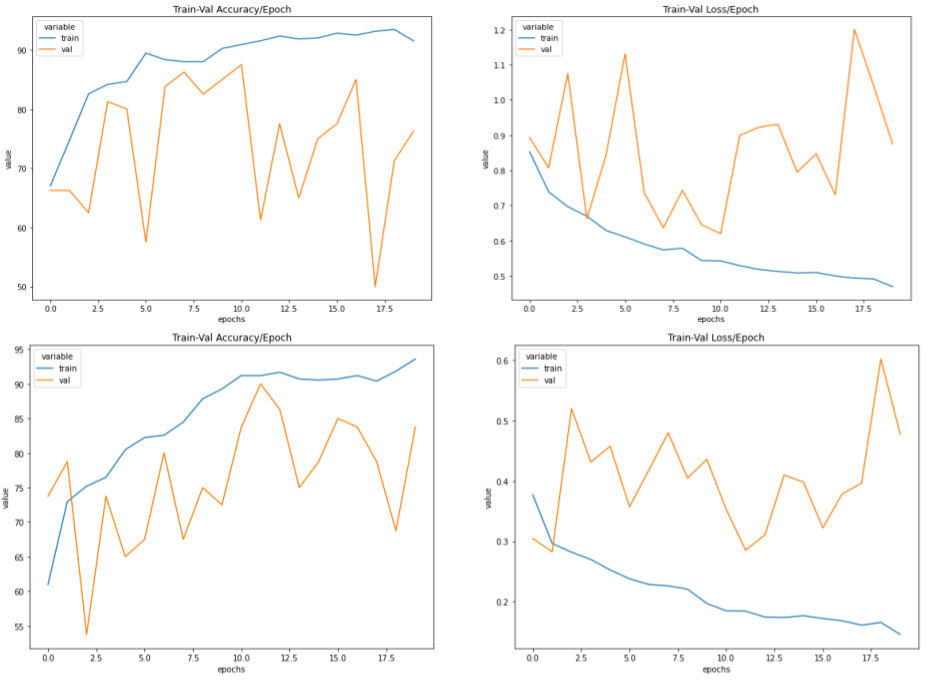}
    \end{adjustbox}
    \label{fig5}
\end{figure}

\subsection{ResNet50 (RCE)}
In this section, we train our model based on ResNet50 without considering transfer learning. The loss function is a focal loss to reduce the imbalanced distribution of sample data. 

\begin{figure}[!htbp]
    \centering
    \begin{adjustbox}{addcode={\begin{minipage}{\width}}{\caption{%
        ResNet50 without pretraining considering FL
        }\end{minipage}},right}
        \includegraphics[width=0.7\textwidth]{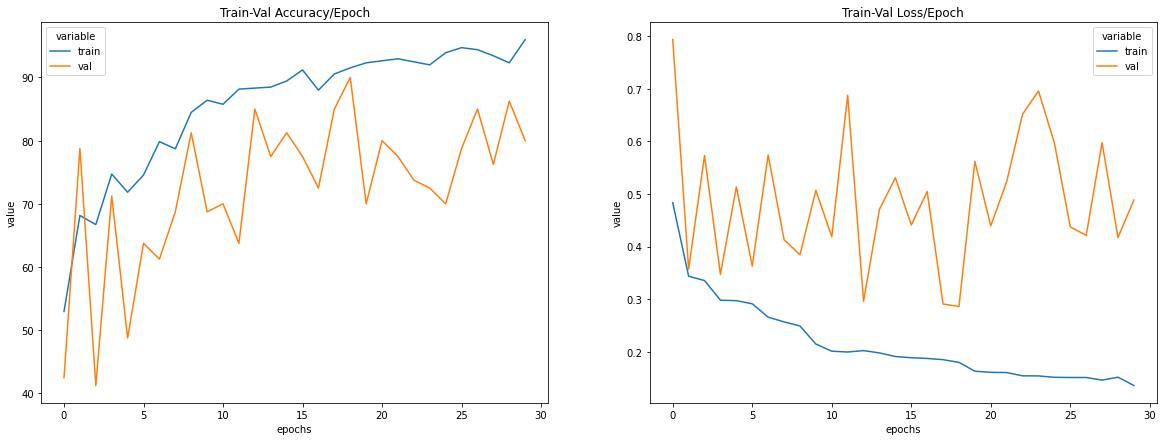}
    \end{adjustbox}
    \label{fig6}
\end{figure}

\subsection{Extra COVID-19 samples}
The limited number of COVID-19 samples is a problematic facts that prevent our model from performing at its best. Thus, at first, we intended to develop a GAN model to generate COVID-19 images. Though, because of a lack of enough time and GPU memory, and the fact that GAN models are very sensitive to the hyperparameters our developed GANs couldn't generate accurate images. Therefore, we decided to add some existing COVID-19 X-ray images on the Internet to our dataset. The results are shown in Table \ref{tab:2}.

\begin{table}[htbp]
  \begin{flushright} 
    \begin{adjustbox}{valign=t}
      \begin{tabular}{lcccccc}
      \toprule
      Models & PRCE & PRCW & PRFL & PDCXCE & PDCXFL & RCE \\
      \midrule
      Tr Acc. & 81.06 & 86.06 & 85.84 & 85.28 & 87.95 & 81.608 \\
      Val Acc. & 65.39 & 73.43 & 75.0 & 76.062 & 73.87 & 69.93 \\
      Tst Acc. & 62.1 & 80.2 & 75.5 & 80.0 & 75.56 & 71.11 \\
      \bottomrule
      \end{tabular}%
    \end{adjustbox}
    \caption{Accuracy results for each proposed model}
    \label{tab:2}%
  \end{flushright}
\end{table}%

\subsection{Discussion}
Based on the results from our implemented models, the highest validation accuracy we've attained falls below 90 percent. This outcome underscores the constraint posed by a limited training dataset. It's our contention that expanding the size of the labeled dataset could enhance the model's accuracy and its capacity to generalize. Nevertheless, acquiring more labeled data presents a time-consuming challenge. Therefore, exploring semi-supervised or self-supervised learning techniques could offer valuable avenues for refining the performance of our model.

\section{Conclusion}
In our study, we explored various strategies for classifying highly imbalanced datasets from Kaggle, focusing on the detection of COVID-19 in lung X-ray images—a critical area of research given the global impact of the pandemic. Our methodology was divided into three principal approaches: utilizing a pre-trained ResNet50 model on ImageNet, employing a pre-trained DenseNet121 on CheXNet, and developing a ResNet50 model from scratch. Our aim was to assess each method's effectiveness to determine the most suitable for analyzing lung X-ray scans. Initial findings indicated that models pre-trained on CheXNet exhibited superior performance on our dataset, surpassing other techniques. Additionally, we observed that implementing focal loss enhanced test and validation set results by addressing the issue of imbalanced input data, despite categorical cross-entropy showing higher training accuracy. A significant challenge was the limited number of COVID-19 samples, which impeded our ability to achieve optimal outcomes. To mitigate this, we incorporated additional COVID-19 samples from another Kaggle challenge. The application of our models, as discussed in section 4, to this augmented dataset led to marked improvements. This underscores the importance of our study in advancing COVID-19 detection, highlighting the potential of machine learning models to improve diagnostic accuracy and, consequently, patient outcomes in the face of this global health crisis.


\def\cprime{$'$} \def\cprime{$'$}
{\color{Brown}
}
\end{document}